\documentclass[prl,superscriptaddress,showpacs]{revtex4}
\usepackage{graphicx,color}
\usepackage{dcolumn}
\usepackage{bm}
\usepackage{amsmath}
\usepackage{amssymb,amsbsy}

\newcommand{\beq}{\begin{equation}}
\newcommand{\eeq}{\end{equation}}
\newcommand{\bea}{\begin{eqnarray}}
\newcommand{\eea}{\end{eqnarray}}
\newcommand{\bec}{\begin{center}}
\newcommand{\enc}{\end{center}}
\newcommand{\bfr}{\begin{flushright}}
\newcommand{\efr}{\end{flushright}}
\newcommand{\alp}{\alpha}

\newcommand{\om}{\omega}
\newcommand{\tom}{\widetilde{\omega}}
\newcommand{\tkap}{\widetilde{\kappa}}
\newcommand{\tgam}{\widetilde{\gamma}}
\newcommand{\tsig}{\widetilde{\sigma}}

\newcommand{\tb}{\widetilde{b}}

\newcommand{\ti}{\widetilde{i}}
\newcommand{\tj}{\widetilde{j}}
\newcommand{\tone}{\widetilde{1}}
\newcommand{\ttwo}{\widetilde{2}}
\newcommand{\tthree}{\widetilde{3}}
\newcommand{\tfour}{\widetilde{4}}
\newcommand{\Om}{\Omega}

\newcommand{\kap}{\kappa}
\newcommand{\gam}{\gamma}
\newcommand{\G}{\Gamma}
\newcommand{\s}{\sigma}

\newcommand{\la}{\langle}
\newcommand{\ra}{\rangle}
\newcommand{\cH}{{\cal H}}

\begin{document}
\title{
Implementation of an impedance-matched $\Lambda$ system 
by dressed-state engineering 
}
\author{Kazuki Koshino}
\affiliation{College of Liberal Arts and Sciences, Tokyo Medical and Dental
University, Ichikawa, Chiba 272-0827, Japan}
\author{Kunihiro Inomata}
\affiliation{RIKEN Center for Emergent Matter Science (CEMS), 2-1 Hirosawa, Wako, 
Saitama 351-0198, Japan}
\author{Tsuyoshi Yamamoto}
\affiliation{RIKEN Center for Emergent Matter Science (CEMS), 2-1 Hirosawa, Wako, 
Saitama 351-0198, Japan}
\affiliation{Smart Energy Research Laboratories, NEC Corporation,  
Tsukuba, Ibaraki 305-8501, Japan}
\author{Yasunobu Nakamura}
\affiliation{RIKEN Center for Emergent Matter Science (CEMS), 2-1 Hirosawa, Wako, 
Saitama 351-0198, Japan}
\affiliation{Research Center for Advanced Science and Technology (RCAST), 
The University of Tokyo, Meguro-ku, Tokyo 153-8904, Japan}
\date{\today}
\begin{abstract}
In one-dimensional optical setups, 
light-matter interaction is drastically enhanced 
by the interference between the incident and scattered fields. 
Particularly, in the impedance-matched $\Lambda$-type three-level systems, 
a single photon deterministically induces the Raman transition 
and switches the electronic state of the system. 
Here we show that such a $\Lambda$ system can be implemented 
by using dressed states of a driven superconducting qubit and a resonator. 
The input microwave photons are perfectly absorbed 
and are down-converted into other frequency modes in the same waveguide. 
The proposed setup is applicable to single-photon detection 
in the microwave domain. 
\end{abstract}
\pacs{
03.67.Lx, 
85.25.Cp, 
42.50.Pq 
}
\maketitle

In one-dimensional optical setups,
radiation from a quantum emitter is guided  
completely to specified one-dimensional propagating modes.
We can realize such setups in a variety of physical systems,
such as optical cavity quantum electrodynamics (QED) systems
using atoms or quantum dots~\cite{oned1,oned2,oned3} 
and circuit QED systems using superconducting qubits~\cite{oned4,oned5,oned6}.
When we apply a field to excite the emitter 
through the one-dimensional mode in these setups,  
the incident field inevitably interferes with the field scattered by the emitter
due to the low dimensionality~\cite{io1}. 
As a result, we can realize unique optical phenomena
that are not achievable in three-dimensional free space. 
A classical example is the complete transmission of 
a resonant field through a two-sided cavity,
in which reflection from the cavity is forbidden
due to the destructive interference between 
the incident field and the cavity emission in the reflection direction. 
Such one-dimensional optical setups 
in which reflection from the emitter is forbidden
are called impedance-matched,
in analogy with properly terminated electric circuits~\cite{imp1,imp2}.
Recently, perfect reflection of 
the incident field by a single emitter has been confirmed in both 
optical cavity QED and circuit QED systems~\cite{oned2,oned5}. 
Here, transmission is forbidden by the destructive interference 
occurring in the transmission direction.

\begin{figure}
\includegraphics[width=85mm]{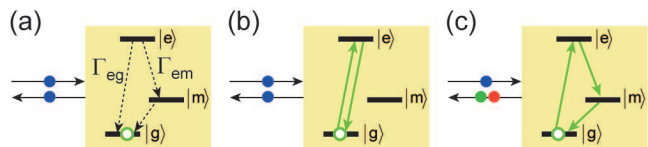}
\caption{\label{fig:Lam}
Interaction between a $\Lambda$ system and 
a photon propagating in a semi-infinite one-dimensional waveguide:
(a)~simple reflection, (b)~elastic scattering, and 
(c)~inelastic scattering. 
$\G_{ij}$ ($i,j=e,m,g$) denotes the radiative decay rate for
$|i\ra\to|j\ra$ transition.
}\end{figure}

In this study, we investigate a three-level $\Lambda$ system
interacting with a semi-infinite one-dimensional field
in a reflection geometry (Fig.~\ref{fig:Lam}).  
We denote the three levels of the $\Lambda$ system by 
$|g\ra$, $|m\ra$ and $|e\ra$ from the lowest.
We assume that $|m\ra$ decays to $|g\ra$ with a finite lifetime
and therefore that the system is in $|g\ra$ when stationary.
When a single photon resonant to the $|g\ra\to|e\ra$ transition is input, 
there are three possible processes:
(a)~simple reflection without exciting the system,
(b)~elastic scattering, inducing the $|g\ra\to|e\ra\to|g\ra$ transitions, 
and (c)~inelastic scattering, inducing the $|g\ra\to|e\ra\to|m\ra\to|g\ra$ transitions. 
Destructive interference occurs here between processes (a) and (b).
In particular, they cancel each other completely
when the two decay rates from the top level $|e\ra$ 
are identical ($\G_{eg}=\G_{em}$)
and the coherence length of the input photon is sufficiently long.
As a result, the input photon is down-converted deterministically, 
inducing the Raman transition in the system
[Fig.~\ref{fig:Lam}(c)]~\cite{dc}.
This is the impedance-matching in the $\Lambda$ system.
The charm of such impedance-matched systems is 
the deterministic electronic dynamics induced by single photons,
which enables novel quantum technologies.
Based on such $\Lambda$ systems,  
single-photon transistors, quantum memories, 
and optical quantum gates have been 
theoretically proposed~\cite{Lam1,Lam2,Lam3,Lam4,Lam5,Lam6}.

In superconducting qubits, 
we use several discrete levels 
formed at the bottom of the anharmonic potential as an artificial atom. 
We usually make the potential symmetric in order to suppress dephasing. 
Then, each eigenstate has a definite parity
and the qubit functions as a ladder-type multilevel system.
We can also make the potential asymmetric,
for example by introducing flux bias in flux qubits. 
The lowest three levels then function as a $\Lambda$ system, 
which have been used to demonstrate, for example,
lasing and cooling of qubits~\cite{las2,las3,las4}.
However, it is difficult to satisfy the impedance-matching condition, 
i.e., identical decay rates from the second excited state, 
in the $\Lambda$ systems thus created.

\begin{figure}
\includegraphics[width=75mm]{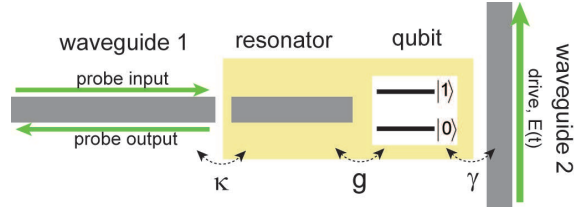}
\caption{\label{fig:Sche}
Schematic of the considered setup.
A qubit is coupled dispersively to a resonator,
which is further coupled to a semi-infinite waveguide (waveguide 1).
The qubit is driven by a microwave field
propagating along another waveguide (waveguide 2).
}\end{figure}

In this study, we propose a practical scheme for implementing
an impedance-matched $\Lambda$ system
by using dressed states of a qubit and a resonator. 
The schematic of the considered setup is shown in Fig.~\ref{fig:Sche}.
A superconducting qubit is coupled to a resonator,
which is further coupled to a semi-infinite waveguide (waveguide~1). 
Through another waveguide (waveguide~2), 
a drive field $E(t)$ is applied to the qubit. 
The qubit functions as a two-level system ($|0\ra$ and $|1\ra$). 
Setting $\hbar=v=1$, where $v$ is the microwave velocity in the waveguides,
the Hamiltonian of the system is 
\bea
\cH(t) &=& \cH_{sys}(t)+\cH_{damp},
\label{eq:H1}
\\
\cH_{sys}(t) &=& \om_q \s^{\dag}\s +\om_r a^{\dag}a 
+g(\s^{\dag}a+a^{\dag}\s)
+\sqrt{\gam}[E(t)\s^{\dag}+E^*(t)\s],
\\
\cH_{damp} &=& \int dk \left[ kb_k^{\dag}b_k
+\sqrt{\kap/2\pi}(a^{\dag}b_k+b_k^{\dag}a)
\right]
+\int dk \left[ kc_k^{\dag}c_k
+\sqrt{\gam/2\pi}(\s^{\dag}c_k+c_k^{\dag}\s)
\right].  
\eea
The meanings of the operators are as follows:
$\s$ and $a$ are the annihilation operators
of the qubit and the resonator, respectively,
and $b_k$ ($c_k$) is the photon annihilation operator
in waveguide~1 (2) with wave number $k$.
The meanings of the parameters are as follows:
$\om_q$ and $\om_r$ are the resonance frequencies of the qubit and resonator, respectively,
$g$ is the qubit-resonator coupling, and 
$\kap$ ($\gam$) is the decay rate of the resonator (qubit) into waveguide 1 (2).
For simplicity, $\gamma$ is assumed to include
the nonradiative decay of the qubit.
We consider the case in which the qubit and the resonator are highly detuned
($|\om_r-\om_q|\gg g$) and are coupled dispersively. 
The drive field is monochromatic, $E(t)=Ee^{-i\om_dt}$,
and is close to the resonance of the qubit. 
We employ the following parameters:
$\om_q/2\pi=5$~GHz, 
$\om_r/2\pi=10$~GHz, 
$g/2\pi=500$~MHz,
$\kap/2\pi=20$~MHz, and  
$\gam/2\pi=1$~MHz.

By switching to the frame rotating at the drive frequency $\om_d$,
the Hamiltonian becomes static.
Then, $\cH_{sys} = (\om_q-\om_d)\s^{\dag}\s+(\om_r-\om_d)a^{\dag}a 
+g(\s^{\dag}a+a^{\dag}\s)+\sqrt{\gam}(E\s^{\dag}+E^*\s)]$.
$\cH_{damp}$ remains unchanged except that
the photon frequency is measured from $\om_d$.
We define the dressed states of the qubit and the resonator
by the eigenstates of $\cH_{sys}$.
We denote them by $|\tj\ra$ and their energies by $\tom_j$ 
($j=1,2,\cdots$) from the lowest.
In the dressed-state basis, the Hamiltonian is rewritten as
\bea
\cH &=& \cH_{sys}+\cH_{damp},
\label{eq:H2}
\\
\cH_{sys} &=& \sum_j \tom_j\tsig_{jj},
\\
\cH_{damp} &=& \int dk \left[ kb_k^{\dag}b_k
+\textstyle{\sum_{i,j}}\sqrt{\tkap_{ij}/2\pi}
(\tsig_{ij}b_k+b_k^{\dag}\tsig_{ji})
\right]
+\int dk \left[ kc_k^{\dag}c_k
+\textstyle{\sum_{i,j}}\sqrt{\tgam_{ij}/2\pi}
(\tsig_{ij}c_k+c_k^{\dag}\tsig_{ji})
\right],
\eea
where $\tsig_{ij}=|\ti\ra\la\tj|$ and
$\tkap_{ij}$ ($\tgam_{ij}$) is the radiative decay rate into waveguide 1 (2)
for $|\ti\ra\to|\tj\ra$ transition.
$\tkap_{ij}$ and $\tgam_{ij}$ are respectively given by
\bea
\tkap_{ij} &=& \kap|\la\ti|a^{\dag}|\tj\ra|^2,
\label{eq:tkap}
\\
\tgam_{ij} &=& \gam|\la\ti|\s^{\dag}|\tj\ra|^2.
\label{eq:tgam} 
\eea
Thus, the time-dependent Hamiltonian 
in the bare-state basis [Eq.~(\ref{eq:H1})]
is transformed into a static one 
in the dressed-state basis [Eq.~(\ref{eq:H2})].

\begin{figure}
\includegraphics[width=70mm]{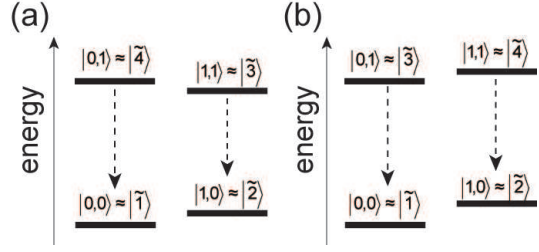}
\caption{\label{fig:LS}
Structure of the four lowest levels of 
the qubit-resonator system for $E=0$:
(a)~Nesting and (b)~unnesting regimes.
The nesting regime is realized when the drive frequency satisfies
$\om_q-3\chi<\om_d<\om_q-\chi$.
Arrows indicate the direction of the cavity decay.
}\end{figure}

For $g=E=0$, the eigenstates of $\cH_{sys}$ are simply 
the product Fock states of the qubit and the resonator, 
$|m,n\ra=|m\ra_{\rm q}|n\ra_{\rm r}$ ($m=0,1$ and $n=0,1,\cdots$).
The qubit-resonator coupling $g$ mixes these states only slightly 
due to the large detuning and brings about dispersive level shifts. 
Within the second-order perturbation, the eigenenergies are given by
\bea
\om_{|0,n\ra} &=& n(\om_r-\om_d+\chi),
\\
\om_{|1,n\ra} &=& \om_q-\om_d-\chi+n(\om_r-\om_d-\chi),
\label{eq:om1n}
\eea
where $\chi=g^2/(\om_r-\om_q)$. 
In this study, we investigate the case in which a weak probe field is 
input from waveguide~1.
Therefore, only the four lowest levels ($|0,0\ra$, $|1,0\ra$, 
$|0,1\ra$, and $|1,1\ra$) are relevant.
Their energy diagrams are shown in Fig.~\ref{fig:LS} for $E=0$.
Due to the dispersive level shifts,
with the proper choice of the drive frequency $\om_d$
($\om_q-3\chi<\om_d<\om_q-\chi$),
the level structure becomes nested, i.e.,
$\om_{|0,0\ra}<\om_{|1,0\ra}<\om_{|1,1\ra}<\om_{|0,1\ra}$ [Fig.~\ref{fig:LS}(a)].
When $\om_d$ is out of this range, the level structure becomes unnested
[Fig.~\ref{fig:LS}(b)]. 
We refer to the former (latter) case
as the nesting (unnesting) regime hereafter.

\begin{figure}
\includegraphics[width=85mm]{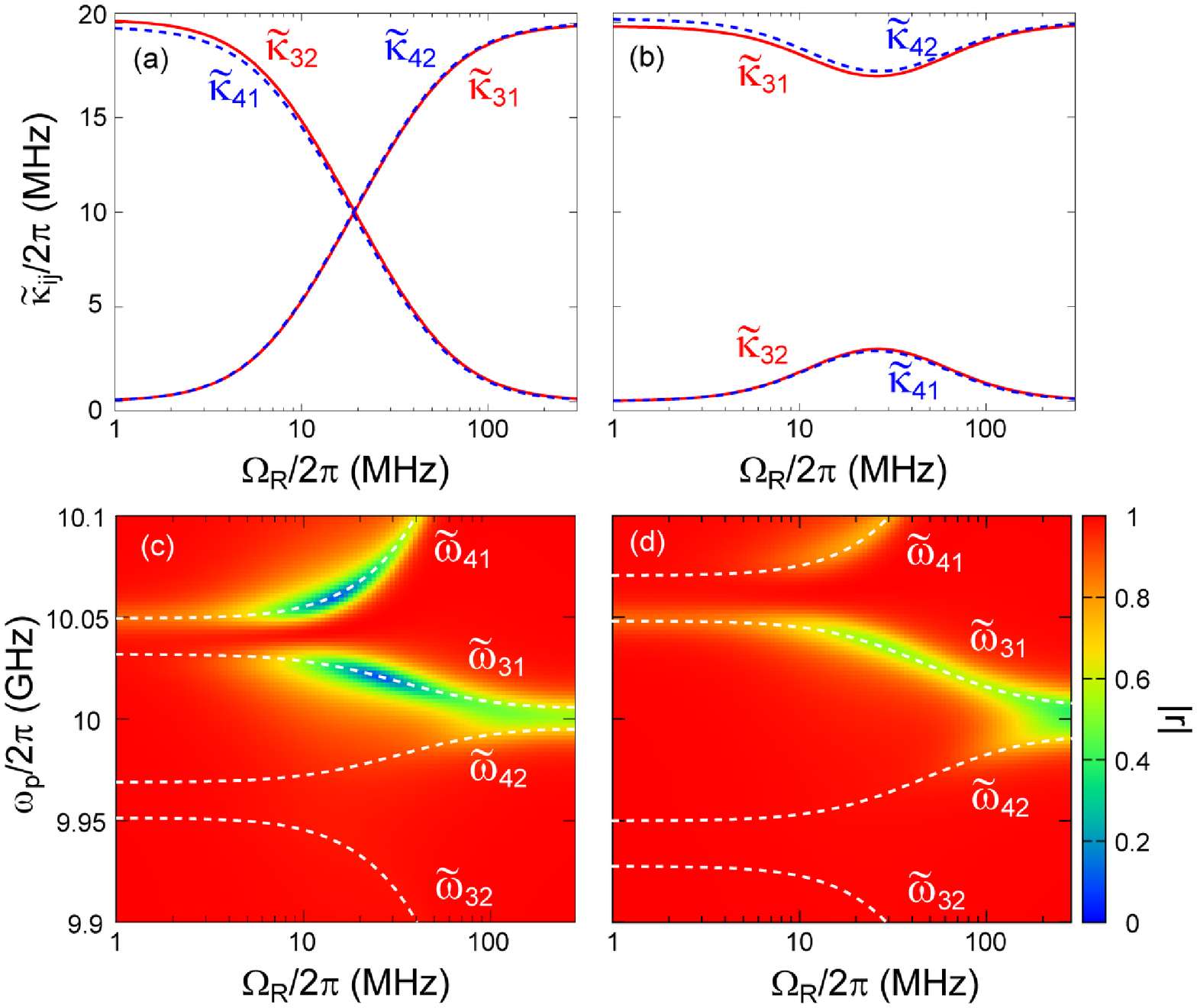}
\caption{\label{fig:4abcd}
(a)~Dependences of $\tkap_{31}$, $\tkap_{32}$,
$\tkap_{41}$ and $\tkap_{42}$ on the drive power. 
The drive frequency is in the nesting regime ($\om_d/2\pi=4.87$~GHz). 
The drive power is expressed 
in terms of the Rabi frequency, $\Om_R=\sqrt{\gam}|E|$. 
The curves for $\tkap_{31}$ and $\tkap_{42}$
($\tkap_{32}$ and $\tkap_{41}$) are mostly overlapping.
(b)~The same plot as (a) in the unnesting regime ($\om_d/2\pi=4.83$~GHz).
(c)~Reflection coefficient $|r|$
as a function of the drive power and the probe frequency
in the nesting regime ($\om_d/2\pi=4.87$~GHz).
(d)~Same plot as (c) in the unnesting regime ($\om_d/2\pi=4.83$~GHz).
In (c) and (d), the probe field is weak ($|F|^2=10^4$~photons/s)
and is in the linear-response regime.
}\end{figure}

Next, we discuss the effects of driving.
The drive field mixes the lower (higher) two levels in Fig.~\ref{fig:LS}
to form dressed states $|\tone\ra$ and $|\ttwo\ra$
($|\tthree\ra$ and $|\tfour\ra$). 
Therefore, neglecting the slight mixing originating in the dispersive coupling, 
dressed states are roughly written as
$|\tone\ra \simeq \cos\alp|0,0\ra-\sin\alp|1,0\ra$,
$|\ttwo\ra \simeq \sin\alp|0,0\ra+\cos\alp|1,0\ra$,
$|\tthree\ra \simeq \cos\beta|0,1\ra-\sin\beta|1,1\ra$, and
$|\tfour\ra \simeq \sin\beta|0,1\ra+\cos\beta|1,1\ra$,
where $\alp$ and $\beta$ 
depend on the frequency $\om_d$ and the power $|E|^2$ of the drive field. 
From Eq.~(\ref{eq:tkap}),
the radiative decay rates $\tkap_{ij}$ into waveguide~1
are given by $\tkap_{31} \simeq \tkap_{42} \simeq \kap\cos^2(\alp-\beta)$,
$\tkap_{32} \simeq \tkap_{41} \simeq \kap\sin^2(\alp-\beta)$, and others vanish.
For weak drive, $(\alp,\beta) \simeq (0,\pi/2)$ and accordingly 
$\tkap_{32}\simeq\tkap_{41}\simeq\kap$ 
in the nesting regime [Fig.~\ref{fig:LS}(a)],
whereas $(\alp,\beta) \simeq (0,0)$ 
and accordingly $\tkap_{31}\simeq\tkap_{42}\simeq\kap$ 
in the unnesting regime [Fig.~\ref{fig:LS}(b)].
In contrast, for strong drive, 
where the Rabi splittings overwhelm the dispersive level shifts,
$(\alp,\beta) \simeq (\pi/4,\pi/4)$ and therefore 
$\tkap_{31}\simeq\tkap_{42}\simeq\kap$ 
in both nesting and unnesting regimes. 
In Fig.~\ref{fig:4abcd}(a) and (b),
$\tkap_{31}$, $\tkap_{32}$, $\tkap_{41}$ and $\tkap_{42}$ 
are evaluated rigorously from Eq.~(\ref{eq:tkap}),
using the Rabi frequency, $\Om_R=\sqrt{\gam}|E|$,
as a measure of the drive power.
We observe that $\tkap_{31} \simeq \tkap_{42}$ and 
$\tkap_{32} \simeq \tkap_{41}$ at any drive power
in accordance with the above discussion. 
Remarkably, inversion of these decay rates occurs
in the nesting regime [Fig.~\ref{fig:4abcd}(a)], 
and the two radiative decay rates from 
$|\tthree\ra$ or $|\tfour\ra$ become identical
with the proper choice of the drive power ($\Om_R/2\pi=19$~MHz).
At this drive power, the qubit-resonator system
functions as an impedance-matched $\Lambda$ system,
with $|g\ra=|\tone\ra$, $|m\ra=|\ttwo\ra$, and 
$|e\ra=|\tthree\ra$ or $|\tfour\ra$.
It is numerically confirmed that 
the lower two levels are mixed only slightly;
$|g\ra\simeq|0,0\ra$ (qubit ground state) 
and $|m\ra\simeq|0,1\ra$ (qubit excited state).
Therefore, the decay rate $\G_{mg}$ is of the qubit origin
and $\G_{mg}\simeq\gam$,
while $\G_{eg}$ and $\G_{em}$ are of the cavity origin
and $\G_{eg}=\G_{em}\simeq\kap/2$.

In the following, we analyze the microwave response 
of this qubit-resonator system 
to a probe field applied through waveguide~1. 
From the Hamiltonian of Eq.~(\ref{eq:H2}), 
the Heisenberg equation for $\tsig_{ij}$ is 
\bea
\frac{d}{dt}\tsig_{ij}&=&i\tom_{ij}\tsig_{ij}
-(\xi^{\kap}_{ij}+\xi^{\gam}_{ij})/2
+i[\zeta^{\kap\dag}_{ij}b_{in}(t)+\zeta^{\gam\dag}_{ij}c_{in}(t)]
-i[b_{in}^{\dag}(t)\zeta^{\kap}_{ji}+c_{in}^{\dag}(t)\zeta^{\gam}_{ji}],
\eea
where $S_{\mu}=\sum_{i,j}\sqrt{\widetilde{\mu}_{ij}}\tsig_{ji}$
($\mu=\kap,\gam$),
$\xi^{\mu}_{ij}=\tsig_{ij}S_{\mu}^{\dag}S_{\mu}
+S_{\mu}^{\dag}S_{\mu}\tsig_{ij}
-2S_{\mu}^{\dag}\tsig_{ij}S_{\mu}$, and
$\zeta^{\mu}_{ij}=[\tsig_{ji},S_{\mu}]$.
The input and output field operators, 
$b_{in}(t)$ and $b_{out}(t)$, are connected by
\beq
b_{out}(t)=b_{in}(t)-iS_{\kap}(t).
\eeq
We assume that a monochromatic probe field
with amplitude $F$ and frequency $\om_p$
is applied from waveguide~1, 
while no probe field is applied from waveguide~2,
i.e., $\la b_{in}(t)\ra=Fe^{-i(\om_p-\om_d)t}$ and $\la c_{in}(t)\ra=0$.
Note that the probe frequency $\om_p$ is measured
from the drive frequency $\om_d$ 
since we are working in the rotating frame.

We define the reflection coefficient
by the ratio of output and input amplitudes, i.e.,
$r=\la b_{out}(t)\ra/\la b_{in}(t)\ra$. 
For weak probe, the system exhibits linear response
and therefore $r$ is independent of the probe power. 
In Fig.~\ref{fig:4abcd}(c) and (d), 
$|r|$ is plotted as a function of the drive power and the probe frequency, 
together with the relevant transition frequencies between dressed states.
Since dissipation (decay into waveguide~2) is negligible here,
the attenuation of $|r|$ results from the inelastic scattering. 
We observe in Fig.~\ref{fig:4abcd}(c) that 
strong suppression of the reflected field amplitude
takes place in the nesting regime as a result of the impedance matching.
The conditions are that 
(i)~the decay rates from $|\tthree\ra$ ($|\tfour\ra$) to $|\tone\ra$ and $|\ttwo\ra$
are identical [$\Om_R/2\pi\simeq 19$~MHz in Fig.~\ref{fig:4abcd}(a)], 
and that (ii)~the probe frequency is tuned to $\tom_{31}$ ($\tom_{41}$). 
Since level $|\tilde{2}\ra$ is almost unoccupied for a weak probe power,
no specific signal appears at $\tom_{32}$ or $\tom_{42}$.
We observe in Fig.~\ref{fig:4abcd}(d) that 
impedance matching never occurs in the unnesting regime.

\begin{figure}
\includegraphics[width=75mm]{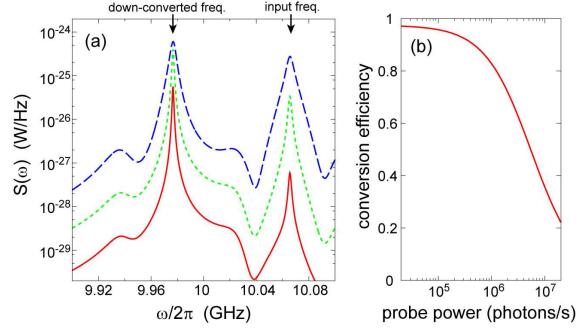}
\caption{\label{fig:pow}
(a)~Power spectrum of the output field
under the impedance-matching condition
($\om_d/2\pi=4.87$~GHz and $\Om_R/2\pi=19$~MHz).
The input probe frequency is tuned to $\tom_{41}$
($2\pi\times 10.066$~GHz),
whereas the dominant peak appears at $\tom_{42}$ ($2\pi\times 9.977$~GHz).
The probe power $|F|^2$ is $10^5$~photons/s (solid), 
$10^6$ (dotted) and $10^7$ (dashed), respectively. 
(b)~Down-conversion efficiency as a function of the probe power.
}\end{figure}

Although the probe amplitude vanishes, 
this does not imply dissipation of the probe power.
Figure~\ref{fig:pow}(a) plots
the power spectrum of the output field,
$S(\om)={\rm Re}\int_0^{\infty}d\tau e^{-i(\om-\om_d)\tau} 
\la \tb_{out}^{\dagger}(t+\tau)\tb_{out}(t)\ra/\pi$,
under the impedance-matching condition.
The input probe is tuned to $\tom_{41}$($=2\pi\times 10.066$~GHz).
However, upon the interaction with the qubit and the resonator, 
the probe field is down-converted nearly completely 
and forms a dominant peak at $\tom_{42}$($=2\pi\times 9.977$~GHz). 
Figure~\ref{fig:pow}(b) plots the down-conversion efficiency,
which is defined by the area of the down-converted peak
normalized by the input power, $|F|^2$.
We observe that, at low probe power, 
most input photons are down-converted.
The conversion efficiency is slightly less than unity
even in the weak-probe limit.
This is due to the qubit-origin decay of level $|\tfour\ra$ into waveguide~2.
The conversion efficiency decreases as the probe power increases.
This is due to saturation of the $\Lambda$ system:
The bottleneck process here is the $|m\ra\to|g\ra$ decay,
the rate of which is approximately $\gam$. 
Therefore, when the input flux exceeds $\gam^{-1}$
($|F|^2\gtrsim 10^6$~photons/s),
$|m\ra$ is populated gradually
and the system becomes insensitive to the probe.

Two final comments are in order. 
(i) For impedance matching, 
achieving the nested energy diagram [Fig.~\ref{fig:LS}(a)] is essential 
and therefore a large dispersive shift $\chi$ is advantageous~\cite{Forn}. 
In this regard, the systems in the so-called straddling regime
are promising, in which the dispersive shift is enhanced 
by the presence of the second excited state of the qubit~\cite{Ino,stra1}. 
Numerical results are qualitatively unchanged 
if we extend the model in this direction. 
(ii)~When a resonant photon with pulse length $\tau$
is input from waveguide~1, 
it induces the $|g\ra\to|e\ra\to|m\ra$ transition
nearly deterministically provided that $\tau\gtrsim\kap^{-1}$~\cite{dc}.
As discussed, $|g\ra$ and $|m\ra$ respectively correspond to
the qubit's ground and excited states.  
Namely, a single photon deterministically excites the qubit.
Therefore, by performing the dispersive 
quantum-nondemolition readout of the qubit~\cite{read2,read3}
within a relatively long qubit lifetime ($\sim\gamma^{-1}$),
we can apply this setup to the detection of single microwave 
photons~\cite{det1,det2,det3}. 
A large dispersive shift is advantageous also in this regard.

In summary, we proposed a circuit-QED implementation 
of an impedance-matched $\Lambda$ system. 
We considered a setup composed of a driven superconducting qubit,
a resonator and a waveguide. 
The lowest four dressed states of the qubit-resonator system,
$|\tone\ra$, $|\ttwo\ra$, $|\tthree\ra$, and $|\tfour\ra$, 
are relevant in this study.
With the proper choice of the drive frequency and intensity,
two radiative decay rates from $|\tthree\ra$ or $|\tfour\ra$ become identical;
the system then functions as an impedance-matched $\Lambda$ system,
where $|g\ra=|\tone\ra$, $|m\ra=|\ttwo\ra$ and $|e\ra=|\tthree\ra$ or $|\tfour\ra$.
When a probe field tuned to the $|g\ra\to|e\ra$ transition 
is applied from the waveguide, 
the probe field loses its coherent amplitude 
and is down-converted nearly completely. 
The present setup is applicable to the detection of single microwave photons.

This work was partly supported by 
the Funding Program for World-Leading Innovative R\&D 
on Science and Technology (FIRST), 
Project for Developing Innovation Systems of MEXT, 
MEXT KAKENHI (Grant Nos. 21102002, 25400417),
SCOPE (111507004),
and National Institute of Information and Communications
Technology (NICT).


\end{document}